# Understanding Electric Vehicle Ownership Using Data Fusion and Spatial Modeling †

Meiyu (Melrose) Pan, Ph.D. **(Corresponding Author)**

National Transportation Research Center

Oak Ridge National Laboratory

1 Bethel Valley Road, Oak Ridge, TN 37830

ORCiD: 0000-0003-1627-448X

Email: panm@ornl.gov

Majbah Uddin, Ph.D.

National Transportation Research Center

Oak Ridge National Laboratory

1 Bethel Valley Road, Oak Ridge, TN 37830

ORCiD: 0000-0001-9925-3881

Email: uddinm@ornl.gov

Hyeonsup Lim, Ph.D.

National Transportation Research Center

Oak Ridge National Laboratory

1 Bethel Valley Road, Oak Ridge, TN 37830

Email: limh@ornl.gov

Declarations of interest: none

† This manuscript has been authored by UT-Battelle, LLC, under contract DE-AC05-00OR22725 with the US Department of Energy (DOE). The US government retains and the publisher, by accepting the article for publication, acknowledges that the US government retains a nonexclusive, paid-up, irrevocable, worldwide license to publish or reproduce the published form of this manuscript, or allow others to do so, for US government purposes. DOE will provide public access to these results of federally sponsored research in accordance with the DOE Public Access Plan (http://energy.gov/downloads/doe-public-access-plan).

# Understanding Electric Vehicle Ownership Using Data Fusion and Spatial Modeling

**Abstract**

The global shift toward electric vehicles (EVs) for climate sustainability lacks comprehensive insights into the impact of the built environment on EV ownership, especially in varying spatial contexts. This study, focusing on New York State, integrates data fusion techniques across diverse datasets to examine the influence of socioeconomic and built environmental factors on EV ownership. The utilization of spatial regression models reveals consistent coefficient values, highlighting the robustness of the results, with the Spatial Lag model better at capturing spatial autocorrelation. Results underscore the significance of charging stations within a 10-mile radius, indicative of a preference for convenient charging options influencing EV ownership decisions. Factors like higher education levels, lower rental populations, and concentrations of older population align with increased EV ownership. Utilizing publicly available data offers a more accessible avenue for understanding EV ownership across regions, complementing traditional survey approaches.

**Keywords**

Electric vehicle; Built environment; Charging stations; Spatial Regression; Public data; Data fusion

# 1. Introduction

The global trend towards the adoption of electric vehicles (EVs) has gained significant momentum, primarily driven by the need to combat climate change and reduce reliance on fossil fuels. Acquiring a comprehensive understanding of the patterns observed in EV ownership can effectively inform the formulation of more people-centric policies and the design of infrastructure aimed at bolstering the widespread uptake of EVs.

Research has shown that land use design can impact climate change mitigation behavior and potentially influence EV purchases (Ford et al., 2018). However, the influence of the built environment on EV ownership has not received adequate attention, and the spatial heterogeneity of this impact remains less explored. Some studies have investigated the impact of the built environment on EV ownership and found that factors such as charger density and road priority, access to restricted traffic zones, positively correlate with the market share of EVs (Wang et al., 2019; R. Zhang et al., 2021). However, a thorough exploration of whether these factors exhibit regional variations is lacking, underscoring the necessity for further research that employs more comprehensive spatial modeling techniques.

Ignoring these regional differences could obscure the true motivations or barriers behind EV ownership. Several studies have found that EV charging stations accelerate EV adoption (Gehrke & Reardon, 2022; Sathaye & Kelley, 2013). Nevertheless, they further raise the question of whether proliferating charging infrastructure universally would truly augment EV ownership. A proper charging network eases EV owner concerns, raising ownership, yet sufficient stations may await more users. Uncertainty exists on if more charging infrastructure can lead to high utilization (Mastoi et al., 2022). It is possible that the infrastructure would mainly benefit existing EV owners and have limited impact in regions with low EV ownership. Therefore, conducting a more nuanced analysis of factors, including their geographical impact, is crucial for developing effective strategies to promote EV uptake in specific regions.

This study endeavors to address the existing research gap by examining the influence of various socioeconomic and built environmental factors on the ownership of EVs, utilizing publicly available data. Currently, stated preference surveys have been widely utilized to understand EV ownership (Jia & Chen, 2021; Massiani, 2014). However, survey approach is less cost-effective when it comes to comprehending the impact of various factors on a larger scale. In response to this challenge, we adopted a data fusion strategy, integrating multiple publicly available datasets, including American Community Survey conducted by US Census Bureau and Smart Location Database published by US Environmental Protection Agency (US Census Bureau, 2021; US Environmental Protection Agency, 2021). By analyzing diverse datasets, we aim to explore the relationships between social and built environment factors and EV ownership.

The findings reported in this study could be used in the formulation of policies and the design of infrastructure aimed at fostering EV ownership. By identifying the socioeconomics and built environment factors that contribute to higher levels of EV ownership and understanding how these relationships vary across different geographical areas, policymakers could tailor promotion strategies and/or infrastructure placement for EV based on the specific needs and characteristics of each region.

# 2. Literature Review

## 2.1 Influencing factors of EV ownership

The literature on factors influencing EV ownership identifies four key domains: demographic, contextual, built environment, and psychological (Gürcan, 2018; Pan et al., 2023; Pevec et al., 2020; Simsekoglu, 2018). Comprehensive reviews have been conducted (Corradi et al., 2023; Singh et al., 2020).

### 2.1.1 Demographic factors

Attention has been devoted to understanding the impact of demographic variables on EV ownership in the last two decades. Factors such as gender, age, education, occupation, and travel patterns have been extensively studied (e.g., Egbue et al., 2017; Javid & Nejat, 2017). Findings indicate that adults, mature individuals, the middle-aged or higher, those with higher education, and professionally employed individuals exhibit a strong inclination towards adopting EVs (Bjerkan et al., 2016; Farkas et al., 2018; Plötz et al., 2014). Furthermore, studies have explored the interplay between income and EV purchase costs, as well as the impact of fuel expenses (Parker et al., 2021; Vega-Perkins et al., 2023). Lower-income demographics exhibit greater sensitivity to both purchase costs and potential fuel savings compared to their higher-income counterparts (Axsen et al., 2015; Hackbarth & Madlener, 2016; Jia & Chen, 2021).

### 2.1.2   Psychological factors

Various psychological factors significantly shape consumers' attitudes towards EVs, including environmental concern, consumer innovativeness, range anxiety, neighborhood effects, motives for car use, and self-assessed knowledge of EVs (Chu et al., 2019; Featherman et al., 2021; Sovacool, 2017; Xia et al., 2022). Among these, environmental concern emerges as the most extensively researched determinant of EV adoption (Austmann & Vigne, 2021; Bai et al., 2020; H. He et al., 2021). Data often collected through surveys and choice models are employed in understanding these factors, while several psychological theories, such as the Unified Theory of Acceptance and Use of Technology (I & II), Theory of Planned Behavior, Technology Acceptance Model, and Innovation Diffusion Theory, provide conceptual frameworks for modeling EV ownership (Ajzen, 1991; Davis, 1985, 1989; Rogers, 2003; Venkatesh et al., 2003, 2012). Leveraging these theories enhances the clarity of data collection paths and improves the interpretability of results (Haustein & Jensen, 2018; Sovacool, 2017).

### 2.1.3   Contextual factors

Studies of contextual factors, encompassing policy and marketing strategies, have underscored the substantial impact of government policy incentives, including preferential tax treatments, exemptions from tolls and parking charges, driving privileges, reduced acquisition and value-added taxes, charging infrastructure incentives, electricity subsidies, bus lane access, road tax exemptions, and fossil fuel taxes, on consumers' purchasing intentions (Lévay et al., 2017; Ouyang et al., 2021; Renaud-Blondeau et al., 2023). Geographical variations have also been explored (Habich-Sobiegalla et al., 2018; van der Steen et al., 2015; Zimm, 2021). For instance, one study compared early adopter characteristics in China and South Korea (Chu et al., 2019). The research revealed that environmental concern played a pivotal role for Chinese early majority adopters, while Korean early adopters were significantly influenced by lower fuel costs and government subsidies.

### 2.1.4   Built environment factors

While sociodemographic factors have garnered more attention, the literature has comparatively given relatively less focus to the impact of built environment characteristics on EV ownership. Nonetheless, studies have demonstrated that the design of compact, walkable neighborhoods can stimulate active travel and the utilization of shared mobility options (Wali et al., 2021). Additionally, land use design has been found to support climate change mitigation efforts (Ford et al., 2018). By strategically designing spaces, such as implementing mixed-use developments and multimodal facilities, urban planners can complement EV adoption by providing alternative transportation modes for shorter trips. Some studies have analyzed spatial characteristics and their influence on EV ownership, considering factors such as fuel economy and the cost of EVs (Bansal et al., 2015; Zhuge et al., 2020). However, built environment factors, such as infrastructure and transit-related aspects, have not been adequately explored. A few studies have focused on the development of parking and charging locations for EV owners (Chen et al., 2013; J. Li et al., 2018). However, many of these studies primarily concentrated on improving system efficiency from the supply side rather than specifically addressing user-oriented improvements.

## 2.2 EV ownership modeling approaches

Numerous researchers have employed discrete choice models, often at an individual or household level, to delve into various vehicle ownership decisions. Various choice models, including logit, mixed logit, and probit, have been utilized (Bailey et al., 2015; S. Y. He, Sun, et al., 2022; Javid & Nejat, 2017; Jia & Chen, 2021). These models often use binary or Likert scales to express ownership or the inclination to own an EV. At their core is the assumption that travelers aim to maximize their utility (Goulias, 2002). The logit model, widely adopted, assumes an Independent, Identically Distributed extreme value error term with a logistic distribution. It produces outcomes similar to probit regression, which employs a standard normal distribution for the error term (Hsiao, 1992). To overcome the limitations of standard logit, mixed logit was proposed, allowing for people's random preference (Train, 2009).

In the realm of understanding behavioral patterns, survey data has traditionally taken center stage, yet the potential of public datasets remains noteworthy. Several studies have dipped into these datasets, often characterized by broader geographical coverage, particularly in applications related to strategic planning for EV infrastructure (S. Y. He, Kuo, et al., 2022; G. Li et al., 2022). Researchers have also utilized public datasets to explore factors like land use density and socioeconomic variables impacting both EV and conventional vehicle ownership. Spatial models, such as Spatial Lag and Spatial Error models, have been applied to analyze these datasets (Chen et al., 2015; X. Liu et al., 2017). With the emergence of data collection initiatives, particularly those focused on public charging stations, there is an opportunity to further investigate charging-related influences and explore new dimensions in the spatial analysis of EV ownership.

## 2.3 Summary of research gaps and contributions

The development of discrete choice models to examine this relationship is hindered by challenges in scalability to broader regions. Specifically, grasping how relationships derived from these models apply to different regions without additional survey efforts might pose challenges. This prompts the question: beyond choice models, what alternative methods exist to explore this relationship? Additionally, how do these influences vary across regions, and is there spatial dependency on EV ownership?

This study seeks to overcome these challenges by leveraging various public data sources to evaluate the impact of socioeconomic and built environment factors on EV ownership on a relatively large scale, specifically all ZIP Code areas within a state. The primary goal is to evaluate the feasibility of employing public data sources for investigating this relationship. To accomplish this, we employ both linear and various spatial regression models, seeking to gauge the consistency of the modeling outcomes. The agreement among the results from these models would affirm the reliability of the relationship between built environment factors and EV ownership identified in the data, implying that public datasets could serve as valuable tools for examining the influencing factors of EV ownership.

Our contribution lies in demonstrating that a spatial data processing and modeling framework utilizing data sources such as Census data and network attributes could help evaluate the impact of socioeconomic and built environment factors on EV ownership. This serves as a proof-of-concept for employing multi-source public data in large-scale EV ownership modeling. Importantly, this framework can offer a practical and complementary approach for public agencies to utilize readily available public data, augmenting traditional survey methods, to understand ways to promote EV ownership.

# 3. Materials and Methodology

## 3.1 Data Sources

This study utilized five publicly available datasets to conduct a comprehensive analysis on EV ownership. Specifically, the datasets were EValuateNY, Smart Location Database (SLD), Alternative Fuels Data Center (AFDC), American Community Survey (ACS), and Center for Neighborhood Technology (CNT).

### 3.1.1    EValuateNY

The first dataset, EValuateNY (NYSERDA, 2023), played a crucial role as it provided the necessary information regarding the target variable that is the EV ownership in New York State at the 5-digit ZIP Code Tabulation Area (ZCTA5) level. In EValuateNY, the EV ownership rate was quantified by calculating the number of electric vehicles registered per 1,000 individuals based on their home location as an ownership indicator. This metric served as a key indicator to assess the extent of electric vehicle adoption across different regions within New York State. Note that the data obtained from EValuateNY were used up to December 2021. Figure 1 illustrates the distribution of EV ownership, with the ZCTA5 zones displayed in white denoting areas without any EV registrations. Given the prevalence of relatively low EV ownership in several regions, we performed a Box-Cox transformation on the target variable. To tackle the challenge of geographical units with zero EV ownership, we mitigated this by adding one to their values.

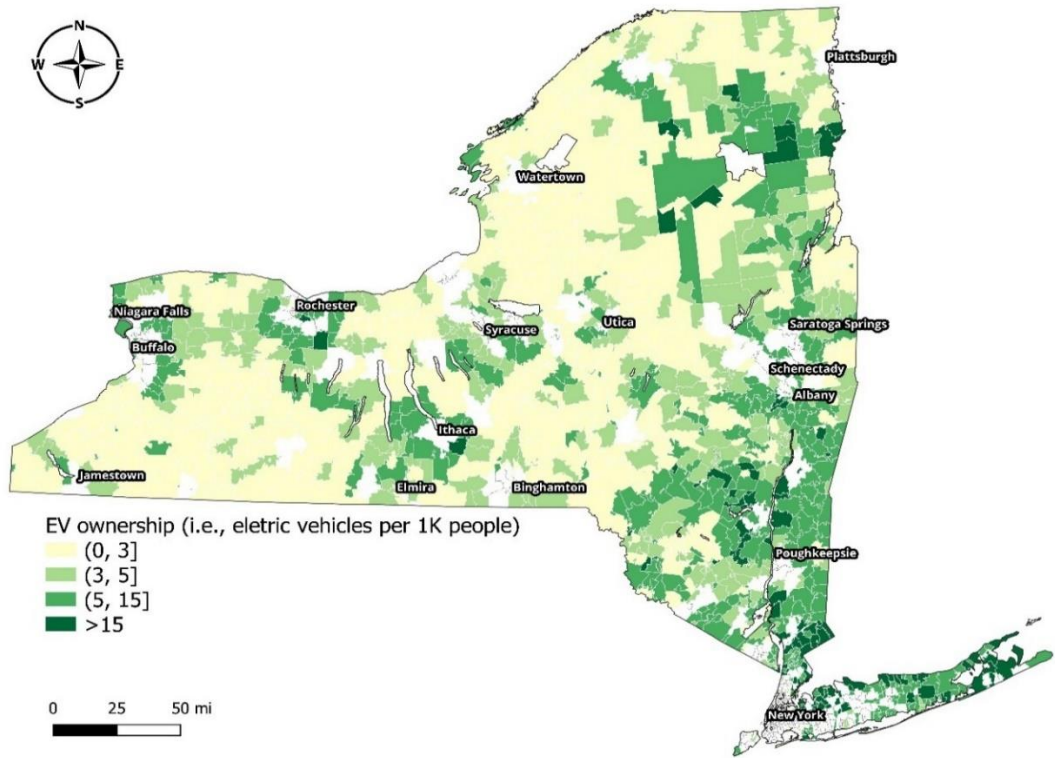

**Figure 1. Distribution of EV ownership in New York State by December 2021 (Source: EValuateNY)**

### 3.1.2    Smart Location Database (SLD)

In order to select the attributes related to the built environment, we adopted the selection criteria proposed by Ewing and Cervero (Ewing & Cervero, 2010). This criteria extends from the three core dimensions, or "3 Ds," of the built environment: density, diversity, and design (Cervero & Kockelman, 1997). The "five Ds" include density, design, diversity, distance to transit, and destination accessibility. Additionally, we considered accessibility to charging stations as another built environment attribute that could potentially influence EV ownership (S. Y. He, Sun, et al., 2022). The reason for utilizing this criteria lies in its comprehensive coverage of factors describing the built environment. Existing literature has demonstrated correlations between factors under these categories, such as residential density, urban classification, and network density, with travel behavior (Bhat et al., 2009; Ding et al., 2021; Saelens & Handy, 2008). This criteria can help us to encompass a broad range of built environment factors, providing a nuanced understanding of their potential impact on EV ownership.

To obtain the built environment data, we utilized the Environmental Protection Agency's (EPA) SLD. The built environment features from SLD were derived from several datasets, such as the 2018 ACS (e.g., vehicle ownership and population), 2018 HERE Maps, and 2020 General Transit Feed Specification (e.g., transit stops). These datasets summarize over 80 attributes for every Census Block Group (CBG) in the United States.

### 3.1.3    Alternative Fuels Data Center (AFDC)

Another data source used is the AFDC, which provides the locations of EV charging stations across the United States (Alternative Fuels Data Center, 2023). Specifically, for this study, we obtained the locations of electric charging stations by ZCTA5 that were operational by the end of 2021. The geographic distribution of the EV charging stations is visually represented in Figure 2. In summary, the built environment features, extracted from both the SLD and AFDC datasets, are outlined in Table 1.

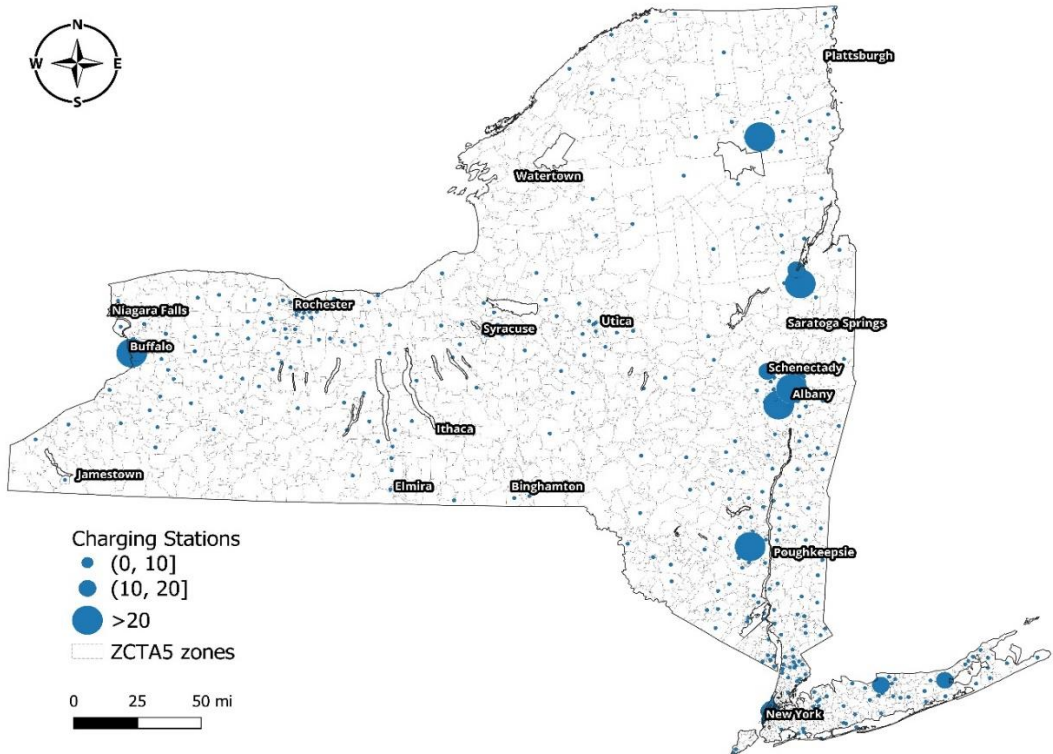

**Figure 2. Distribution of EV charging stations in New York State in 2021 (Source: AFDC)**

**Table 1. Built-environment attributes explored in this study**

| Dimensions | Description | Variables | Unit | Sources |
|---|---|---|---|---|
| Density | The variables of interest per unit of area | Population density | People per acre | SLD |
| | | Residential density | Household units per acre | SLD |
| | | Employment density | Jobs per acre | SLD |
| | | Retail employment density | Jobs per acre | SLD |
| | | Office employment density | Jobs per acre | SLD |
| Design | The characteristics of the street network inside an area | Low speed network density (i.e., miles of low-speed road segments (e.g., road segments having speeds between 31 and | Miles of road segments per square mile | SLD |
| | | 40 mph) based on HERE speed category per square mile | | |
| | | High speed network density (i.e., miles of high-speed road segments (e.g., road segments having speeds are 55 mph or higher) based on HERE speed category per square mile) | Miles of links per square mile | SLD |
| Diversity | The number of different land uses in a fixed area and the represented degree | Employment entropy (i.e., the diversity of retail, office, industrial, service, and entertainment jobs) | No unit | SLD |
| | | Employment and household entropy | No unit | SLD |
| Distance to transit | The level of transit service at the residences or workplaces | Distance to nearest transit stop (from the population-weighted centroid) | Meters | SLD |
| | | Transit frequency | Transit per hour per square mile | SLD |
| Destination accessibility | Ease of access to a certain location | Number of jobs within 45 minutes transit travel time | Jobs | SLD |
| | | Number of jobs within 45 minutes auto travel time | Jobs | SLD |
| | | Number of charging stations | Charging stations within a ZCTA5 area | AFDC |
| | | | Charging stations within 5, 10, 25, 50, 75, and 100 miles from the ZCTA5 centroid | AFDC (derived) |

### 3.1.4   American Community Survey (ACS)

Regarding socioeconomic attributes, we first explored ACS data. The ACS is an annual nationwide survey designed to provide reliable and timely social, economic, housing, and demographic data for various geographical entities. From the ACS, we gathered variables identified in the literature as potentially influential on EV ownership, such as the proportion of older populations (i.e., aged 65+), education level, gender, the proportion of rental housing, and the proportion of White population. In particular, we utilized the ACS 5-year estimate, spanning the period from 2017 to 2021.

### 3.1.5   Center for Neighborhood Technology (CNT)

The latest CNT data used in this study was derived from the ACS and the Longitudinal Employer-Household Dynamics data in 2019 (Center for Neighborhood Technology, 2022). Specifically, this study utilized the CNT's Housing + Transportation (H+T) index, which incorporates two main components: home burden and transportation burden. The home burden reflects the proportion of housing costs in relation to household income, while the transportation burden reflects the proportion of transportation costs in relation to household income. This comprehensive dataset has gained widespread adoption among transportation planning agencies, facilitating the identification of suitable locations for development initiatives, such as investments in new transit systems or the construction of affordable housing (M. Zhang, 2022). Conventional metrics such as median income from the ACS might possess limitations in fully capturing the disposable income disparities across multiple regions within New York State. In contrast, the H+T index offers a more comprehensive assessment of the relative affordability of distinct places. A summary of these attributes explored in this study is provided in Table 2.

**Table 2. Socioeconomic attributes explored in this study**

| Variable | Description | Source | Unit |
|---|---|---|---|
| % Older | The proportion of populations age 65+ | ACS | ZCTA5 |
| % White | The proportion of populations that are White | ACS | ZCTA5 |
| % Rent | The proportion of populations that rent | ACS | ZCTA5 |
| % Bachelor | The proportion of populations with bachelor's degree of higher | ACS | ZCTA5 |
| % Zero-vehicle household | The proportion of household with zero vehicle | ACS | ZCTA5 |
| Median Income | Household annual median income | ACS | ZCTA5 |
| Home burden | The proportion of home cost on income | CNT | Census Tract |
| Transportation burden | The proportion of transportation cost on income | CNT | Census Tract |

## 3.2 Spatial Lag Model, Spatial Error Model, and Geographically Weighted Regression

In this study, we utilize three widely used spatial regression models, which are the Spatial Lag model, the Spatial Error model, and the Geographically Weighted Regression (GWR) model.

In a Spatial Lag model, the formulation is given by Equation 1:

$$y_i = \beta_0 + x_i\beta + \rho W_i y_i + \varepsilon_i \qquad \text{Equation 1}$$

Where $y_i$ is the $i^{th}$ ZCTA5 location of the target variable, which in this context is EV ownership measured by electric vehicles per 1,000 people. The $x$ variables denote the explanatory variables, encompassing built environment and socioeconomic factors. $\beta_0$ is the estimated coefficient of the intercept, $\beta$ represent the estimated coefficients of the independent variables $x$, $\rho$ signifies the spatial lag parameter, $W_i$ is the spatial weight vector, and $\varepsilon$ denotes the vector of error terms.

A Spatial Error model is specified as Equation 2:

$$y_i = \beta_0 + x_i\beta + u_i, u_i = \rho W_i u_i + \varepsilon_i \qquad \text{Equation 2}$$

Comparing these two models reveals some similarities, but there are distinctions worth noting. In Spatial Lag models, spatial autocorrelation is modeled by a linear relation between the response variable $y_i$ and the associated spatially lagged variable $W_i y_i$. On the other hand, Spatial Error models incorporate spatial autocorrelation through an error term $u_i$ and the associated spatially lagged error term $W_i u_i$.

A significant spatial lag term may indicate strong spatial dependence, suggesting that EV ownership in the neighborhood could influence the EV ownership in the ZCTA5 under consideration. A significant spatial error term suggests spatial autocorrelation in errors, potentially due to omitted key explanatory variables. Both of these regression types are considered "global" because their coefficient estimates exhibit spatial stationarity, allowing a single model to be applied uniformly across different areas of interest.

Third, the study adopts the Geographically Weighted Regression (GWR) model. Previous research indicates that GWR sometimes offers improved model fit compared to traditional "global" regression, assuming relationships are constant over space (Fotheringham, Brunsdon, and Charlton 2002). However, the extent to which this model can effectively explain spatial effects is not yet definitively established, as heightened model fit may also suggest overfitting. GWR introduces local regression, enabling coefficients to vary at each observation. The GWR model extends the conventional Ordinary Least Square (OLS) model by incorporating a geographical location parameter, expressed as Equation 3:

$$y_i = \beta_{i0} + \sum_{k=1}^{p} \beta_{ik} X_{ik} + \varepsilon_i, where \; i = 1,2,...,n$$

<div align="right">Equation 3</div>

Here, *p* denotes the number of independent variables. The distinction of GWR from global models lies in the spatially varying coefficient $\beta_{ik}$, representing the coefficient t of $x_k$ at ZCTA5 location *i*.

In the context of spatial lag and spatial error models, our study employed the widely used Queen contiguity weight matrix (Fotheringham & Rogerson, 2008; Kasu & Chi, 2019; Tong et al., 2013). This matrix remains constant across all observations and is binary in nature, signifying neighbors based on shared borders or edges. Specifically, for each pair of observations, the entry in the weight matrix is set to 1 if they share a common boundary, and 0 otherwise. The Queen contiguity weight matrix identifies neighbors by virtue of entities sharing a border or a point. It characterizes a global spatial relationship without accounting for local variations. On the other hand, GWR takes a different approach. Instead of relying on a binary weight matrix as seen in Spatial Lag and Spatial Error models, GWR employs a continuous and smoothly decaying weights matrix with distance that adapts locally for each observation (Aghayari et al., 2017; Charlton & Fotheringham, 2009; Runhua Xiao et al., 2022). This distinction allows GWR to capture spatially varying coefficients, with the weight matrix changing dynamically across locations.

Regarding model assumptions, all three models align with the generalized linear regression assumptions, including linearity, independence of explanatory variables, homoscedasticity (constant variance of errors), and the assumption of normality for errors. Concerning spatial dependence, the Spatial Lag model assumes spatial lag dependence in the response variable. The Spatial Error model assumes spatial autocorrelation in errors. Both Spatial Lag model and Spatial Error model assume a constant spatial relationship for all locations. They provide a global measure of spatial dependence. Meanwhile, GWR assumes spatially varying coefficients for each explanatory variable.

## 4. Research Design
### 4.1 Data processing

The target variable was obtained based on the geographical classification of ZIP Codes. The US Census Bureau states that, in most cases, the ZCTA code corresponds to the ZIP Code for a given area (US Census Bureau, 2022). Consequently, for the sake of simplicity, ZIP Code was directly assumed to be equivalent to the ZCTA5 unit. Subsequently, the remaining variables used in the study were converted into the ZCTA5 geographical units. However, it is important to acknowledge a potential limitation stemming from the merging of different geographic hierarchies. The ZCTA5 unit can extend beyond the boundaries of a Census Tract, leading to misclassification bias in the study's results. Nonetheless, in order to maintain coherence with the target variable, the independent variables measured at the Census Tract level were transformed into the ZCTA5 units.

This conversion from Census Tract to ZCTA5 employed the ZCTA-Census Tract crosswalk data (US Census Bureau, 2010). The crosswalk data provides, for each ZCTA5, the percentage of populations from a Census Tract overlapped onto the ZCTA5. Each Census Tract was assigned to the ZCTA5 containing the highest percentage of its population. This approach addresses limitations associated with distance-based matching methods, often overlooking population mismatches (Ong & Miller, 2005). Similar population-weighted techniques for matching ZCTA5 with Census Tracts or counties have been employed in other spatial analyses (Bor et al., 2023; Daly et al., 2018).

To establish the lowest population weight percentage allowing a ZCTA5 to be matched to a Census Tract, a sensitivity analysis was conducted. As depicted in Figure 3, the number of successfully matched ZCTA5 decreases as the population weight percentage increases, following an almost linear relationship. It is important to note that there is no universally correct value for this parameter. To maintain a certain sample size, the chosen value was determined as 20%.

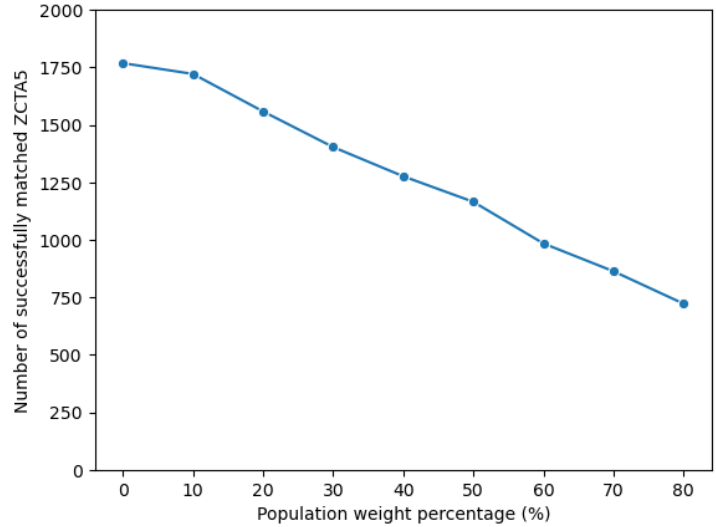

**Figure 3. The relationship between population weight percentage and number of successfully matched ZCTA5**

### 4.2 Descriptive analysis

A total of 1422 ZCTA5 units' data were included in the final dataset for modeling, following the exclusion of regions with missing values for either the target or independent variables. Note that a relatively larger proportion of geographical units were excluded in the Long Island area. To assess multicollinearity, a variance inflation factor (VIF) analysis was performed. The VIF serves to quantify the potential linear dependence among variables, with a VIF value exceeding 10 indicating the presence of multicollinearity.

To maintain uniformity in interpretation, all four models employ an identical set of variables. The rationale behind excluding other attributes is their lack of significance in any of the models. The finalized model exclusively incorporates variables outlined in Table 3. All VIF values in this dataset remained below 2, suggesting the absence of multicollinearity. The independent variables were standardized to facilitate modeling and ensure comparability across different variables. Furthermore, the target variable underwent a Box-Cox transformation.

**Table 3. Summary of the variables used for modeling**

| Variable | Description | Source | Year | Original Geographical Unit | Variable Type | Mean (Standard Deviation) | VIF |
|---|---|---|---|---|---|---|---|
| % Older | The proportion of populations age 65+ | ACS | 2021 | ZCTA5 | Independent (socioeconomic) | 21.0% (9.2%) | 1.1 |
| % Bachelor | The proportion of populations with Bachelor's degree or higher | ACS | 2021 | ZCTA5 | Independent (socioeconomic) | 31.8% (18.5%) | 1.4 |
| % Rent | The proportion of renter- | ACS | 2021 | ZCTA5 | Independent (socioeconomic) | 22.1% (19.5%) | 1.3 |

| | occupied housing units | | | | | | |
| Low speed network density | Miles of low-speed road links per square mile | SLD | 2018 | CBG | Independent (built environment) | 7.3 (5.3) | 1.6 |
| Transit accessible jobs | Number of jobs within 45 minutes transit travel time | SLD | 2020 | CBG | Independent (built environment) | 129750.4 (166480.1) | 1.5 |
| Mixed land uses | Employment and household entropy | SLD | 2020 | CBG | Independent (built environment) | 0.5 (0.1) | 1.1 |
| Home burden | The proportion of home cost on income | CNT | 2019 | Census Tract | Independent (built environment) | 28.0% (7.1%) | 1.4 |
| Charging stations | Number of charging stations | AFDC | 2021 | ZIP Code | Independent (built environment) | 0.8 (2.5) | 1.1 |
| EV ownership | Electric vehicles ownership per 1K people | EValuateNY | 2021 | ZIP Code | Target | 5.6 (6.4) | - |

### 4.3 Model comparison criteria

Moran's I is a useful metric for examining the spatial clustering of model residuals. The Moran's I statistic spans the range from -1, indicative of dispersion, to 1, signifying clustering. The positive or negative sign of the value discerns positive or negative spatial autocorrelation, respectively. Specifically, a positive Moran's I implies a tendency for similar values to co-occur in proximity, while a negative value suggests a dissimilarity in nearby observations.

Within the framework of a regression model, Moran's I plays a crucial role in investigating the spatial autocorrelation of errors between observations. The emergence of a statistically significant p-value for Moran's I can serve as an indicator of a departure from the assumption of independence among model errors. This spatial dependence may provide insights into previously unexplored spatial patterns or overlooked local influences.

Consequently, in this study, we examine the Moran's I value derived from the residuals of each model to discern which models more adeptly capture the spatial association between socio-economic and built environment variables and EV ownership. To facilitate a comprehensive comparison, we also analyze the results of the Ordinary Least Squares (OLS) model alongside the spatial regression models.

In addition to the assessment of spatial dependency, the evaluation of model fit is conducted through established statistical measures, including adjusted R-squared, Akaike Information Criterion (AIC), and Log-Likelihood. To further evaluate the model's performance, a critical examination is undertaken to discern the presence of overfitting. A 5-fold cross-validation is executed as a methodical approach, exposing the model to different subsets of the dataset iteratively. The resultant average training and testing errors serve as pivotal indicators. Specifically, a low training error with a high testing error raises concern, potentially signaling the presence of overfitting. Consequently, this consideration indicates the potential unreliability of coefficients.

## 5. Results

### 5.1 Impact of distance to charging station range

Relying solely on counting the number of charging stations within a ZCTA5 area may not be sufficient to gauge the overall charging infrastructure needs. To address this concern, we conducted additional estimations of the number of charging stations located at varying distances from the ZCTA5 centroid, including 5, 10, 25, 50, 75, and 100 miles. This assessment offers insights into the distances people are likely to consider when deciding to use a charging station.

We calculated the number of charging stations within various distances from the centroid of each ZCTA5 area. The considered distances were 5 miles, 10 miles, 25 miles, 50 miles, and 100 miles. In this analysis, we substituted the original charging station variable (i.e., counts within ZCTA5) with counts within different geographical units, including 5-mile radius, 10-mile radius, 25-mile radius, 50-mile radius, and 100-mile radius, while holding the other independent variables constant.

To determine the most influential distance range of charging stations on EV ownership, we used AIC as a measure of model fit. The results of the four models are analyzed together. Lower AIC values indicate better model fit, signifying a more significant role for the specific range of charging stations. The results from four models were collectively analyzed. We illustrated the relationship between AIC and the different ranges of the charging station variable in Figure 4. The results revealed that the number of charging stations within a 10-mile radius had the strongest correlation with EV ownership across all models, albeit with subtle changes. Consequently, we selected the number of charging stations within 10 miles from the centroid of each ZCTA5 area for inclusion in the final model. This choice ensures that the model captures the most relevant information about the impact of charging station availability on EV ownership.

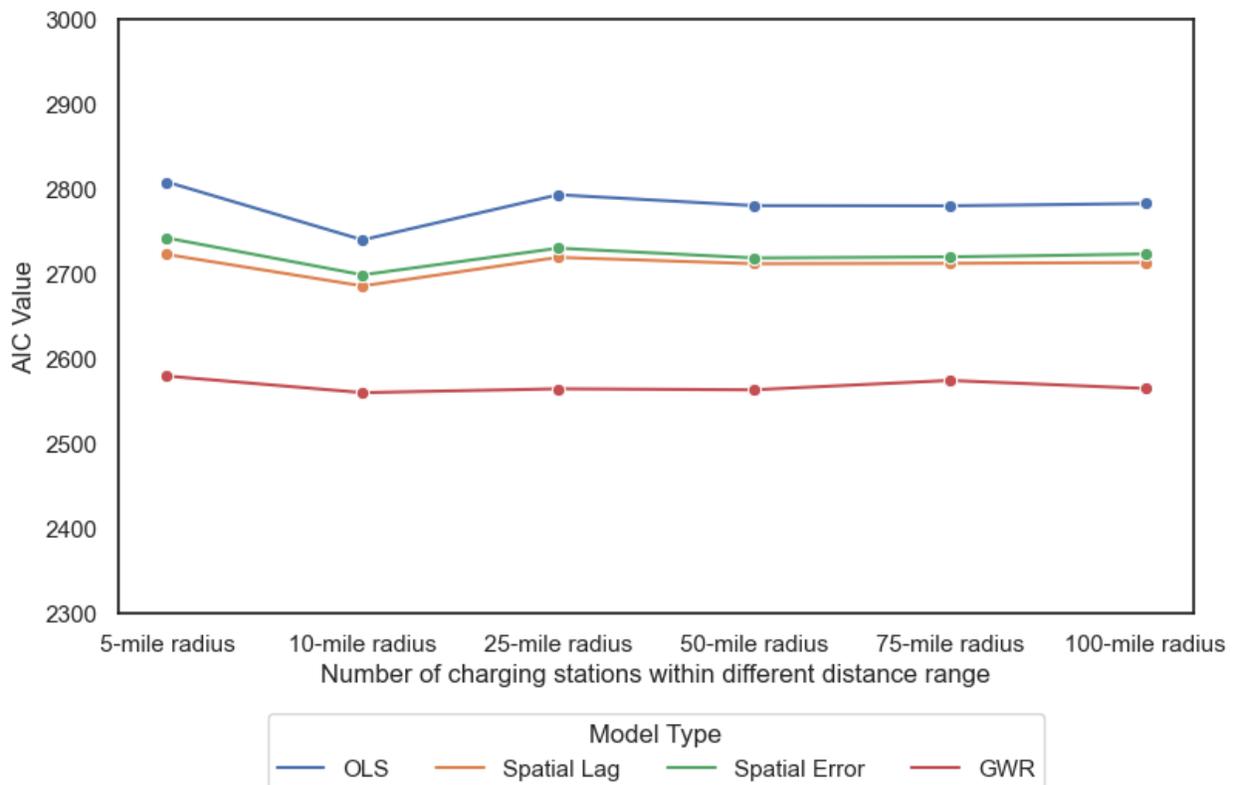

**Figure 4. MGWR AIC by different charging station variables**

### 5.2 Regression results

A summary of the modeling results is presented in Table 4, with coefficients of built environment and socioeconomic variables, along with their standard errors, depicted Figure 5 and Figure 6 respectively.

Upon examining coefficients across the four models, consistent signs are observed for all variables. Except for transit accessible jobs, built environment variables—such as low-speed network density, employment and household entropy, and the number of charging stations—demonstrate a positive influence on EV ownership. The relatively high coefficient for the number of charging stations suggests a notable impact compared to other built environment variables. Conversely, the proportion of rental families exhibits a negative influence on EV ownership. Although low-speed network density positively influences EV ownership, it lacks significance in the Spatial Lag or Spatial Error models. Variables such as the share of individuals with a Bachelor's degree, older populations, and housing burden demonstrate a positive influence, with educational background standing out with a relatively higher impact compared to other socioeconomic variables. A comparison between Figure 5 and Figure 6 reveals a relatively higher standard error for built environment factors compared to socioeconomic variables. This indicates that the current modeling framework may not adequately capture the impact of built environment factors.

Regarding spatial autocorrelation, the Moran's I value derived from the OLS model suggests the presence of spatial autocorrelation in EV ownership. In simpler terms, the basic linear model falls short of fully explaining the variance in EV ownership. Notably, the Moran's I value for the Spatial Lag model is 0.0028, indicating no significant deviation from zero, signifying an absence of spatial autocorrelation in model residuals. This suggests that the Spatial Lag model excels in capturing spatial dependencies among EV ownership by incorporating information from adjacent ZCTA5 areas. In contrast, both the OLS model and Spatial Error model appear less adept at capturing such spatial dependencies.

When comparing model fit, the GWR and Spatial Lag models exhibit relatively higher Adjusted R-Squared values and Log-Likelihoods, along with lower AIC values, suggesting better fits. The Spatial Lag model, however, shows relatively smaller standard errors for most explanatory variables compared to other models. GWR appears to outperform Spatial Lag, with a lower AIC and higher Log-Likelihood.

Since superior model fit may raise concerns about overfitting, a 5-fold cross-validation is conducted, revealing average training and testing errors. None of the models exhibit clear signs of overfitting. Although GWR demonstrates better goodness-of-fit and lower training error than the Spatial Lag model, its slightly higher test error suggests that the Spatial Lag model might be less sensitive to overfitting.

**Table 4. Modeling coefficients, standard error, and summary**

|  |  | OLS | | GWR | | Spatial Lag | | Spatial Error | |
| --- | --- | --- | --- | --- | --- | --- | --- | --- | --- |
|  |  | Coef. | Std. Error | Coef. | Std. Error | Coef. | Std. Error | Coef. | Std. Error |
| Variable | Intercept | 0.06 | 0.11 | 0.22 | 0.13 | 0.00 | 0.10 | 0.16 | 0.11 |
|  | % older people | 0.54** | 0.16 | 0.53 | 0.19 | 0.47*** | 0.15 | 0.45** | 0.16 |
|  | % Bachelor degree of higher | 1.54*** | 0.11 | 1.40 | 0.14 | 1.32*** | 0.11 | 1.41*** | 0.12 |
|  | % Rent | -0.51*** | 0.12 | -0.49 | 0.15 | -0.37*** | 0.12 | -0.39*** | 0.13 |
|  | Low speed network density | 0.30* | 0.15 | 0.25 | 0.18 | 0.23 | 0.14 | 0.23 | 0.16 |
|  | Employment and household entropy | 0.31* | 0.15 | 0.28 | 0.19 | 0.22 | 0.15 | 0.29* | 0.15 |
|  | Transit accessible jobs | -0.12* | 0.05 | -0.11 | 0.06 | -0.09* | 0.05 | -0.09 | 0.06 |
|  | Housing cost/income | 0.95*** | 0.17 | 0.76 | 0.21 | 0.77*** | 0.16 | 0.87*** | 0.18 |
|  | Charging stations | 0.66*** | 0.08 | 0.68 | 0.09 | 0.49*** | 0.08 | 0.62*** | 0.09 |
| Spatial Dependency | Moran I |  | 0.44*** |  | 0.077*** | **0.0028 (*p*=0.429)** |  |  | 0.131*** |
| Goodness-of-fit | Adjusted R2 |  | 0.37 |  | 0.42 |  | 0.41 |  | 0.38 |
|  | AIC |  | 2739.00 |  | 2641.29 |  | 2685.18 |  | 2698.26 |
|  | Log-Likelihood |  | -1360.70 |  | -1298.33 |  | -1332.59 |  | -1340.13 |
| Presence of Overfitting | Training MAE (5-fold CV) |  | 0.457 |  | 0.436 |  | 0.444 |  | 0.461 |
|  | Testing MAE (5-fold CV) |  | 0.481 |  | 0.478 |  | 0.472 |  | 0.476 |

*$p \leq 0.05$  **$p \leq 0.01$  ***$p \leq 0.001$
GWR coefficients do not have global p-value

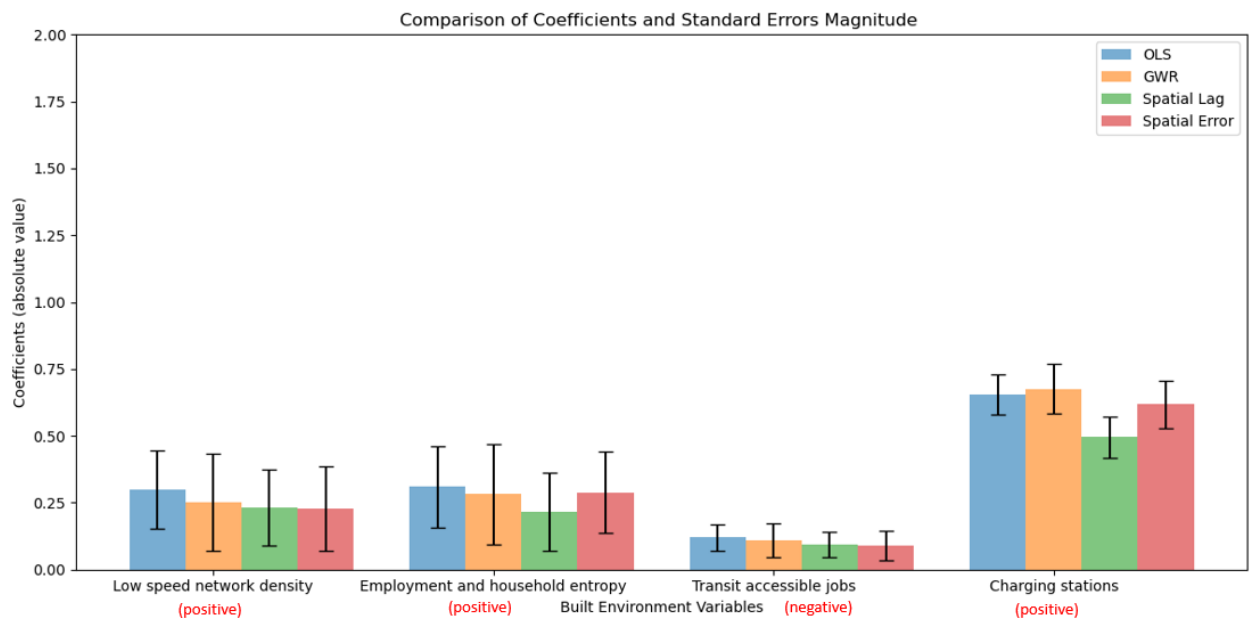

**Figure 5. Coefficients and standard errors of the built environment variables**

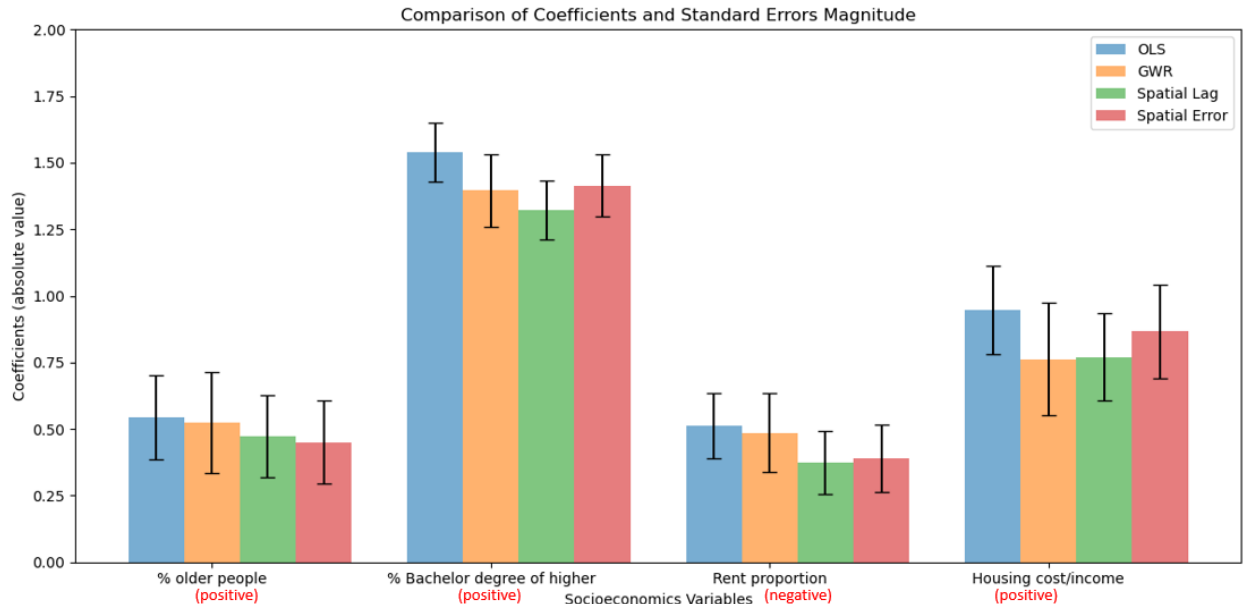

**Figure 6. Coefficients and standard errors of the socioeconomic variables**

## 6. Discussions

### 6.1 Model comparison

Upon reviewing the coefficients of the four models, it becomes apparent that there is relatively small variation among them. This consistency implies a robust relationship between the built environment, socioeconomic factors, and EV ownership. A spatial dependency check highlights the Spatial Lag model's superior ability to capture spatial autocorrelation. Additionally, the Spatial Lag model exhibits lower susceptibility to overfitting.

The relatively higher performance of the Spatial Lag model can be attributed to several factors. A comparison of the model forms of Spatial Lag and Spatial Error Models, extensively discussed in the literature, provides valuable insights. The Spatial Lag regression model incorporates dependent variables in an area with connections to other associated areas, while the Spatial Error regression model accounts for the dependency of error values in an area with errors in other associated areas. Essentially, the Spatial Error model is deemed more suitable when unobserved factors influencing the target variable exist. Our findings suggest that the EV ownership pattern aligns more closely with a Spatial Lag model, where the EV ownership of adjacent zones could influence the EV ownership of a given zone. A comparison of the target variable and its spatial lag, weighted by adjacent neighbors in Figure 7, reveals similarities in their forms. The spatial lag value similarly reflects regions with relatively high or low EV ownership. The Spatial Error model's underperformance in this study implies that unobserved factors might not be significant, and the current set of explanatory variables likely account for a substantial portion of the variance in EV ownership. In practical terms, this finding could be supported by social influence impact, where people are likely to be influenced by those around them to buy EV (Axsen et al., 2013; Pettifor et al., 2017). While this phenomenon is commonly studied at the individual level via surveys, our study extends this understanding to a broader geographical level.

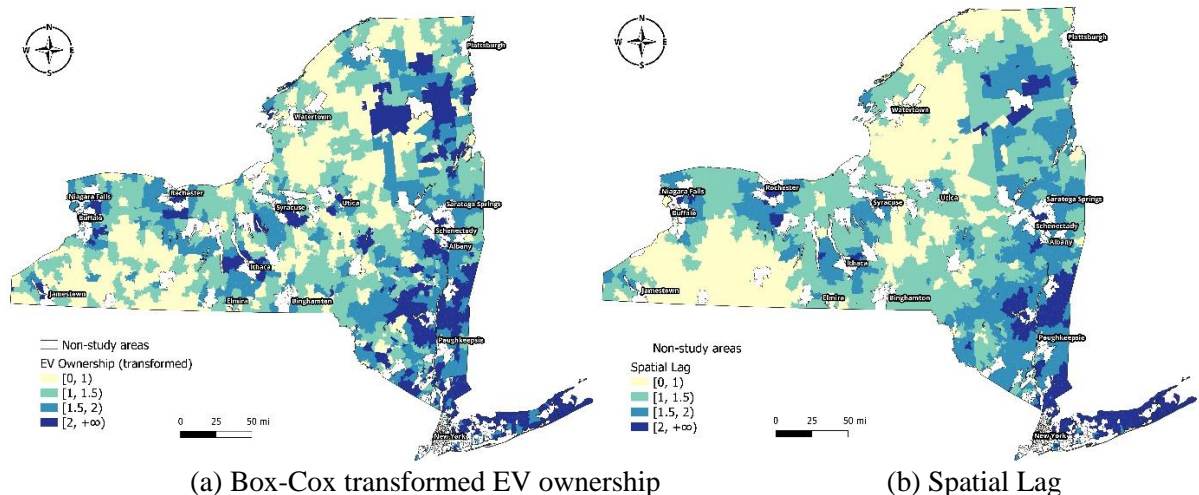

(a) Box-Cox transformed EV ownership      (b) Spatial Lag

**Figure 7. Comparison of target variable and its spatial lag**

Concerning the GWR model, this method entails estimating local models for each spatial location, introducing potential increased model complexity. In our case, where local variations are not evidently pronounced, justifying this complexity may be challenging and could result in poorer model performance. This is evident in GWR's slightly higher susceptibility to overfitting, as indicated by a slightly higher test error compared to the Spatial Lag model. It is worth noting that GWR could potentially benefit from a larger dataset, such as panel data across years for each observation (i.e., ZCTA5 or other zones), to better demonstrate its superiority (Yu, 2010; Yu et al., 2021).

### 6.2 Interpretations of key findings

The Spatial Lag model, alongside other models, consistently indicates a positive association between higher education and increased EV ownership. This relationship can be explained by several interconnected factors. Firstly, education significantly influences income levels (Jia & Chen, 2021; Sovacool et al., 2019). Higher education often provides access to better-paying job opportunities, enabling individuals to afford electric vehicles. Moreover, education correlates with heightened awareness of environmental issues, specifically the impact of conventional vehicles on climate change (Bansal et al., 2015; Egbue & Long, 2012). This environmental consciousness may motivate individuals with higher education to adopt eco-friendly transportation alternatives. Beyond economic and environmental considerations, electric vehicle ownership can symbolize an individual's personal values, particularly those related to environmental awareness (Heffner et al., 2007). We also observed that higher concentrations of older populations correlate with increased EV ownership rates, likely due to their association with higher incomes. Additionally, previous mobility analysis in New York State suggests that older individuals undertake shorter trips on average (Y. Liu et al., 2022; Pan et al., 2023), reducing range anxiety and making EVs more appealing. Some individuals within this demographic group may exhibit consistent adoption of EV. Future research could delve deeper into understanding which groups or regions within the older population might express greater interest in purchasing EVs compared to others. Such insights have the potential to inform strategic initiatives that facilitate the broader integration of EVs into diverse demographic segments.

The share of rental populations showed a negative association with EV ownership, aligning with the fact that about 80% of all EV charging occurs at home (Ge et al., 2021). This implies that a lack of easily accessible charging stations poses a barrier for renters to adopt EVs. While a dedicated home charging place appears to be a necessary condition for EV ownership, the presence of public charging infrastructure could also play a role. The relatively significant coefficient of EV charging stations within a 10-mile radius from a ZCTA5 could reflect people's preferences for the charging range, potentially influencing their decision to own an EV. This range often includes workplaces, retail locations, and transit facilities where public charging stations are frequently utilized (Ai et al., 2018; Borlaug et al., 2023). While most charging occurs at home, having accessible public charging within this radius provides added convenience, potentially enhancing the appeal of EV ownership. This alternative option could contribute to a higher likelihood of adoption. This finding also aligns with research indicating that individuals are more willing to detour to a charging station if the detour is less than ten minutes (Philipsen et al., 2017; Sun et al., 2016).

Current literature indicates that the built environment influences people's vehicle ownership and mode choice; for instance, a high degree of mixed land use may shorten commuting origin-destination distances, reducing the likelihood of owning multiple vehicles (Ding et al., 2018; Zegras, 2010). However, in this study, mixed land use is not found to be significant. Other built environment variables such as low speed network density appear to have relatively higher standard error than socioeconomics variables, indicating that their impact might not be captured well using the existing modeling framework. The Spatial Lag model indicates that the availability of employment opportunities reachable through transit systems may act as inhibiting factors for EV ownership. Individuals may perceive transit-connected job options as convenient and cost-effective, reducing the perceived need for personal vehicle ownership, including EVs, as an alternative mode of transportation (Langbroek et al., 2018).

In summary, our study contributes to the existing literature by introducing a data-driven modeling framework for studying EV ownership. The practical utility of this framework lies in offering public agencies an alternative data collection and modeling perspective, allowing them to understand the factors influencing people's decisions to own an EV without exclusively relying on surveys. While we used New York State as our study region, the data analyzed, such as AFDC and ACS data, is not exclusive to New York State but is readily available in many other states as well. This approach has the potential to save time and effort for public agencies, providing them with a broader regional insight.

### 6.3 Limitations and future studies

There are a few limitations that need to be considered in this study. Firstly, the availability of data on EV ownership at more commonly used geographical units, such as Census Block Groups or Census Tracts, was constrained by the data format of the vehicle registration datasets. Although efforts were made to match Census Tracts with ZCTA5, there were certain regions that had to be excluded from the modeling due to a low level of overlap. This means that the analysis may not fully capture the variations in EV ownership across all geographic areas. Secondly, while the study examined a wide range of built environment characteristics, only a few of them were found to significantly contribute to the model's performance, and consequently, only those were included in the final model. Moreover, despite the primary use of linear regression-based models in this study, the relatively high standard error suggests the possibility of employing nonlinear relationships to better capture the impact of built environment variables. Temporal features such as transit frequency, though potentially influential on EV ownership, weren't extensively explored. Future studies could expand the geographical scope by considering multiple states or regions and explore a broader set of built environment variables to gain a more comprehensive understanding of the factors influencing EV ownership. Additionally, including socioeconomic factors such as proportion of advanced degrees, ownership of more than two vehicles, single-family house prevalence, and average commute distance could be valuable. Yet, these factors might exhibit multicollinearity; for instance, households with multiple vehicles were found strongly correlated with median income and transportation burden. Principal Component Analysis could potentially facilitate forming variable groups, enhancing model input comprehensiveness and efficiency. Regarding the modeling approach, while GWR does not exhibit superior performance compared to the Spatial Lag model, the introduction of panel data could potentially enhance its efficacy. However, the primary built environment SLD data, crucial for this study, has not been compiled for multiple years, restricting the ability to implement a panel data model in the current research. Despite the emphasis on spatial regression models in this study, it is important to acknowledge the possibility of exploring other statistical or machine learning models. Diversifying the modeling approach could offer valuable insights and contribute to a more comprehensive understanding of the relationships inherent in the data. This consideration opens up opportunities to leverage alternative methodologies that might uncover patterns and relationships that spatial regression models alone may not capture.

# 7. Conclusions

This study explores the impact of social and built environment factors on electric vehicle (EV) ownership, utilizing five publicly available datasets: EValuateNY, Smart Location Database (SLD), Alternative Fuels Data Center (AFDC), American Community Survey (ACS), and Center for Neighborhood Technology (CNT). Spatial regression models, including Spatial Lag, Spatial Error, and Geographically Weighted Regression, along with Ordinary Least Squares, are employed to investigate these relationships. The analysis reveals consistent coefficient values across different models, enhancing confidence in the robustness of the findings.

The Spatial Lag model emerges as particularly noteworthy, outperforming other models and suggesting that EV ownership in adjacent zones could significantly influence the EV ownership within a given zone. Furthermore, the proximity of charging stations within a 10-mile radius from a ZCTA5 is revealed to reflect individuals' preferences for convenient charging alternatives, thereby impacting their decision to own an EV. Noteworthy correlations also emerge, indicating that higher education levels, lower rental populations, and concentrations of older populations are associated with increased EV ownership.

This coherence in findings underscores the validity of the models, emphasizing that public data sources serve as effective tools for elucidating the influence of built environment and socioeconomic factors on EV ownership. Leveraging public data presents a more accessible alternative for analyzing these factors across a broader region, potentially complementing traditional survey methods that may be susceptible to variations in stated preferences and actual behavior. The study thus contributes to understanding of the factors influencing EV ownership, shedding light on the practicality and reliability of utilizing publicly available data for such analyses.

**Declaration of generative AI in scientific writing**
During the preparation of this work the authors used Chat GPT-3.5 Model to refine certain language aspects. After using this tool/service, the authors reviewed and edited the content as needed and take full responsibility for the content of the publication.

# Acknowledgement

The authors would like to acknowledge the support of the New York State Department of Transportation (NYSDOT). The opinions, findings, and conclusions in this paper are those of the authors and not necessarily those of the NYSDOT.